\documentclass[usenatbib]{mnras}
\usepackage{epsfig}
\usepackage{amsmath}
\usepackage{graphicx}
\usepackage{array}
\usepackage{textcomp}
\usepackage{amssymb}
\usepackage{caption}
\usepackage{subcaption}
\def\mnras {{\sc MNRAS}}
\def\apj {ApJ}
\def\apjs {ApJS}
\def\apjl {ApJL}

\def\pasp {PASP}
\def\pasj {PASJ}
\def\aap {A\&A} 

\def\apss{Ap\&SS}

\title[Passive Spiral Galaxies]
      {A Photometrically and Spectroscopically Confirmed Population of Passive Spiral Galaxies}
\author[A.\ Fraser-McKelvie et al.]
       {Amelia Fraser-McKelvie$^{1,2}$\thanks{amelia.mckelvie@monash.edu}, 
       Michael J. I. Brown$^{1,2}$, 
        Kevin A. Pimbblet$^{1,2,3}$, \and Tim Dolley$^{1,2}$, Jacob P. Crossett$^{1,2}$, Nicolas J. Bonne$^{1,2,4}$.
        \vspace*{1mm}\\
        $^{1}$ School of Physics and Astronomy, Monash University, Clayton, Victoria 3800, Australia\\
        $^{2}$ Monash Centre for Astrophysics (MoCA), Monash University, Clayton, Victoria 3800, Australia\\
	$^{3}$ E. A. Milne Centre for Astrophysics, University of Hull, Cottingham Road, Kingston-upon-Hull, HU6 7RX, UK\\
	$^{4}$ Institute of Cosmology and Gravitation, University of Portsmouth, Dennis Sciama Building, Burnaby Road, Portsmouth PO1 3FX, UK\\
	}

\begin{document}
\maketitle
\begin{abstract}
We have identified a population of passive spiral galaxies from photometry and integral field spectroscopy.
We selected $z<0.035$ spiral galaxies that have $WISE$ colours consistent with little mid-infrared emission from warm dust.
Matched aperture photometry of 51 spiral galaxies in ultraviolet, optical and mid-infrared show these galaxies have colours consistent with passive galaxies.
Six galaxies form a spectroscopic pilot study and were observed using the Wide-Field Spectrograph (WiFeS) to check for signs of nebular emission from star formation. We see no evidence of substantial nebular emission found in previous red spiral samples. 
These six galaxies possess absorption-line spectra with 4000\AA\ breaks consistent with an average luminosity-weighted age of 2.3 Gyr.
Our photometric and IFU spectroscopic observations confirm the existence of a population of local passive spiral galaxies, implying that transformation into early-type morphologies is not required for the quenching of star formation.

\end{abstract}
 

\section{Introduction}
 The correlation between galaxy morphology and star formation rate is well established \citep[e.g.,][]{Tully82}, and this has been reinforced by Sloan Digital Sky Survey (SDSS) studies of the relationship between galaxy colour bimodality and morphology \citep[e.g.,][]{Baldry04}. However, these results can lead to the misinterpretation that all spiral galaxies are star forming \citep[e.g.,][]{Faber07}. 
This is not the case, as the existence of anaemic disk galaxies with red optical colours associated with little or no star formation has been known for decades \citep[e.g.,][and references therein]{vandenBergh76, Poggianti99, Bundy10, Masters10}, though just how passive these galaxies are when examined in ultraviolet (UV) or infrared (IR) wavelengths is unclear \citep[e.g.,][]{Cortese12}. Given there exist mechanisms for quenching star formation that act with little disturbance to morphology of a galaxy, for example, ram pressure stripping and  strangulation \citep[e.g.][]{Bekki02, Goto03, Bundy10, Tojeiro13}, it is odd we do not yet have a consensus on passive spiral properties as a population. 

Confusingly, the terms `red' and `passive' spiral are often used interchangeably in the literature to describe a spiral galaxy that is optically red. Optically red galaxies can be passive, but they can also have low rates of star formation or suffer from dust-obscured star formation. Spectroscopy typically reveals red spiral galaxies have nebular line emission resulting from low levels of star formation \citep[e.g.][]{Ishigaki07}. In this letter we will define a `red spiral' as a galaxy with red optical colours, even if UV and mid-IR photometry show evidence of star formation.
  We will use the term `passive spiral' to describe galaxies that also possess red optical colours, but lack evidence of star formation in multi-wavelength photometry and spectroscopy. 

The selection criteria employed to find and classify red (and potentially passive) spiral samples are critical; an optical colour cut or the inclusion of edge-on spirals often results in red galaxies with high levels of dust obscuration or small (but significant) star formation \citep[e.g.][]{Gallazzi09, Wolf09, Rowlands12, Dolley14}. 
Moving beyond optical wavelengths can improve sample selection and characterisation, and an IR colour cut can differentiate between dusty star forming galaxies and truly passive populations. The anaemic spirals of \citet{vandenBergh76}, for example, show a range of mid-IR colour. \citet{Bundy10}, used an $R-J$ colour cut to define a $0.2<z<1.2$ passive spiral sample, although this sample lacks spectroscopic confirmation.  
Ideally, photometry should be used in tandem with spectroscopic observations to determine whether truly passive galaxies are being observed.
Unfortunately, this may not be possible with low redshift galaxy surveys using single-fibre spectroscopy, as the small fibre size generally captures just the light of a galaxy bulge, missing the potentially star-forming outskirts \citep[e.g.][]{Ishigaki07}.

Given these limitations, there have been some spread of conclusions on red spiral star formation rates (SFRs) in the literature. While passive spiral candidates have been found in the IR \citep{Bundy10}, 
optically-defined samples of red spiral galaxies often have H$\alpha$ emission \citep[e.g.,][]{Goto03, Ishigaki07, Masters10}, which may be attributed to star formation, AGNs or LINER activity. 
The results of these studies lead us to question whether passive spirals can even exist in the local Universe.

With this in mind, the aim of this letter is to demonstrate conclusively whether passive disk galaxies exist in the local Universe. We have compiled an IR selected sample of passive spiral galaxy candidates, and searched for evidence of star formation using UV, optical and mid-IR photometry. We have then searched for nebular emission and measured luminosity weighted ages for these galaxies using IFU spectra.
 This letter is organised as follows: in Sec.~\ref{sample} we define our IR-selected passive spiral galaxy candidates, in Sec.~\ref{photometry} we use multi-wavelength photometry to confirm whether they are truly passive. In Sec.~\ref{observations1} we describe a pilot sample of six passive spiral galaxy candidates with integral field spectroscopy (IFS), and search for optical emission lines from these galaxies. Throughout this letter we use AB magnitudes and a flat $\Lambda$CDM cosmology, with $\Omega_{m}=0.3$, $\Omega_{\Lambda}=0.7$ and $H_{0}=70 \textrm{km}~\textrm{s}^{-1}~\textrm{Mpc}^{-1}$.
 \section{Sample}
 \label{sample}
\begin{figure*}
 \centering
\includegraphics[width=5.5in, trim={0 2.5cm 0 0},clip]{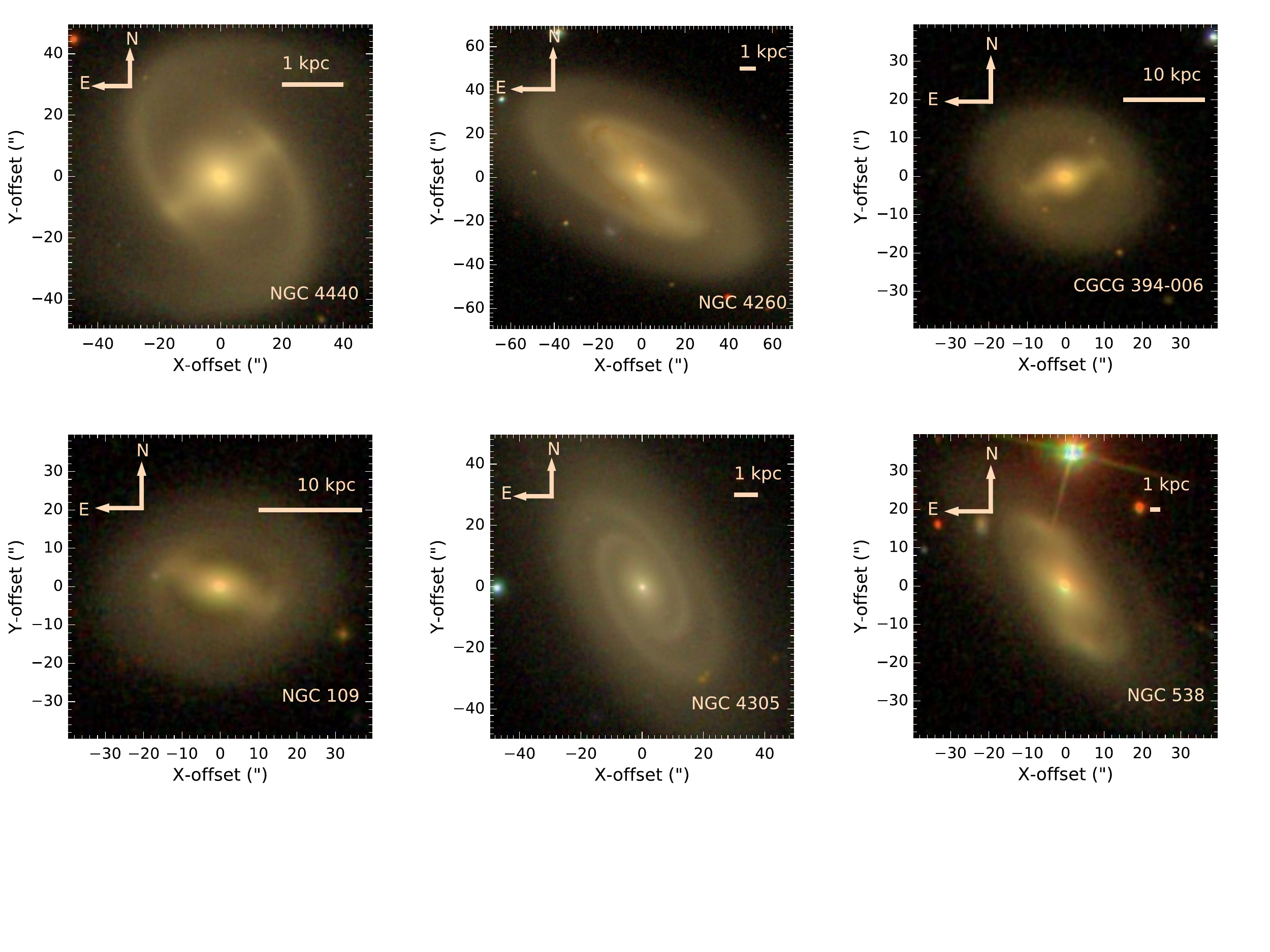}
  \caption{SDSS colour images of some of the passive spiral galaxy candidates in our sample. CGCG 394-006 and NGC 538 are also part of the spectroscopic sample observed for this work using the WiFeS IFS.}
  \label{SDSS_images}
\end{figure*}

 We define two samples of passive spiral galaxy candidates, which partially overlap. We define a photometric sample of 51 galaxies that are located within the SDSS DR10 footprint, which have SDSS imaging to confirm morphologies and optical colours, some examples of which are shown in Fig.~\ref{SDSS_images}. A pilot spectroscopic sample of six galaxies have declinations accessible to the ANU 2.3m telescope at Siding Spring Observatory, and in some instances only Digitised Sky Survey (DSS) photographic plate scans were available to visually confirm morphological type.
 
For both samples, we select galaxies from the parent catalogue of \citet{Bonne15}, which is an all-sky sample of 13,325 local Universe galaxies drawn from the 2MASS Extended Source Catalogue. 
 This catalogue has a redshift and morphological completeness of 99\% down to $K=12.59$, with the majority of morphologies coming from the PGC catalogue \citep{Paturel03}. For further details we refer the reader to \citet{Bonne15}. We select galaxies with $1<\textrm{T-type}<8$, and to search for the most passive galaxies we make the colour cut $K-W3> -2.73$, where $K$ and $W3$ are 2MASS $K$-band and \textit{WISE} 12$\mu$m absolute magnitudes respectively. This cut corresponds to galaxies with a specific star formation rate (sSFR) $<10^{-11.7}\textrm{yr}^{-1}$ (Dolley et al., in prep) based on the $W3$ SFR calibration of \citet{Cluver14} and NASA Sloan Atlas\footnote{\url{www.nsatlas.org}} stellar masses. We classify any galaxies with an sSFR lower than this value as passive spiral candidates. 
 The morphological T-types from \citet{Bonne15} were found to be partially contaminated by lenticular and elliptical galaxies on visual inspection, so for our photometric sample we include only galaxies within the SDSS DR10 imaging regions. For both samples, we classify red spirals by eye, rejecting any edge on galaxies, those without obvious spiral arms, and two galaxies that appear optically blue, likely due to errors in 2MASS or \textit{WISE} photometry. The quality of the SDSS images makes spiral identification straightforward, although the photographic plate images are more ambiguous. 
 
 There are a total of 51 galaxies in the photometric sample in the range $0.002<z<0.035$, with $r$-band magnitudes $14.14<r<11.42$, average stellar mass $5\times10^{10}\textrm{M}\odot$, and SDSS, \textit{WISE}, and frequently \textsc{galex} photometry.
The spectroscopic sample have mean $z=0.02$, median stellar mass of $5.3\times10^{10}\textrm{M}\odot$ and an average SDSS exponential fit radius of 12$''$. The declinations of these six galaxies prevent them from all having SDSS or \textsc{galex} coverage.

 \section{Photometric Analysis}
 \label{photometry}
 We perform UV, optical and mid-IR matched-aperture photometry on our photometric sample of passive galaxy candidates to search for signs of star formation.
 For each galaxy, the aperture is an ellipse that encloses the galaxy light down to a threshold of 25 $\textrm{mag}~\textrm{arcsec}^{-2}$ in the SDSS $r$ band. We measure photometry in SDSS $u,g,r,i,z$ and \textsc{galex} $NUV$ and $FUV$ bands from images provided by the NASA Sloan Atlas, the results of which are presented in Table~\ref{phot_table}. These images have already been background subtracted, though we still use 10 background regions of the same size as the galaxy aperture and mask obvious UV and optical sources within the galaxy aperture. Mid-IR photometry was measured from the \textit{All-WISE} Image Atlas in the 3.4, 4.6, 12 and 22 $\mu$m passbands (bands $W1$-$W4$) using the same background subtraction method as for the UV and optical bands. The point spread function at all wavelengths is much smaller than the angular size of the galaxies, so the seeing in different passbands is not an issue.
A Milky Way foreground dust correction was performed using the dust maps of \citet{Schlafly11} for SDSS and \citet{Yuan13}  for the UV using a reddening coefficient of $R=3.1$.

\begin{figure*}
 \centering
  \includegraphics[width=7in]{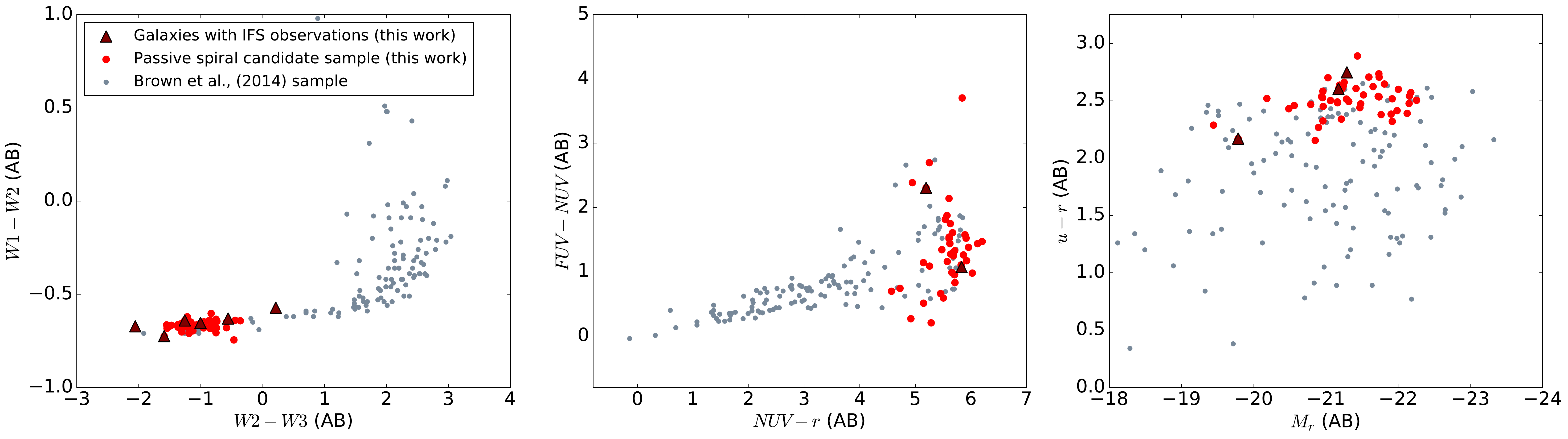}
  \caption{Colour-colour and colour-magnitude photometry plots of passive spiral galaxy candidates in the mid-IR, UV and optical, with the comparison galaxy sample of \citet{Brown14}. In all three panels, our sample lie in the regions usually occupied by passive elliptical galaxies.}
  \label{phot_plots}
\end{figure*}

In Fig.~\ref{phot_plots} we present colour-colour and colour-magnitude plots of our sample. The left panel shows the mid-IR colour-colour diagram of passive spiral galaxy candidates in our photometric sample, the six galaxies with WiFeS IFS observations, and for comparison, the sample of nearby galaxies from \citet{Brown14}, with accurate multi-wavelength photometry spanning a broad range of galaxy types and SFRs. The passive spiral candidate sample clearly sits in the passive region of the diagram, where we would usually expect ellipticals to lie. This is not surprising since these galaxies are selected from mid-IR photometry, but confirms these galaxies are consistent with passive galaxies at mid-IR wavelengths. 
We compare the $WISE$ photometry of our passive spiral candidates with archival photometry of red spiral samples from the literature. Our galaxies have a typical $W2-W3$ colour of -1, with only one galaxy with $W2-W3>0$. The 364 galaxies with $z<0.035$ in the \citet{Masters10} sample have typical colour of $W2-W3\sim1.5$, with just one galaxy with $W2-W3<0$. The \citet{vandenBergh76} anaemic spirals span a range of colour, from $-1<W2-W3<2$, some of which will be truly passive galaxies, and some with star formation.  We conclude that overall, our sample has far less mid-IR emission from warm dust than previous samples of red spiral galaxies.

The centre panel of Fig.~\ref{phot_plots} is a UV colour-colour diagram. Unfortunately, only two of the passive spiral galaxy candidates with WiFeS spectroscopy have \textsc{galex} $NUV$ and $FUV$ photometry, although our entire photometric sample lies in the passive region of the diagram. 
\citet{Cortese12} found that 94\% of the \citet{Masters10} red spiral sample had $NUV-r<4$  (the typical value assumed for actively star forming galaxies), while all of our passive spiral sample are redder than this value, with typical $NUV-r\sim5.5$, consistent with an old stellar population \citep[e.g.][and references therein]{Crossett14}.

In the right panel of Fig.~\ref{phot_plots} we present an optical colour-magnitude diagram of the passive spiral galaxy candidates, and the three galaxies with WiFeS spectroscopy and SDSS photometry. All of the passive spiral galaxy candidates are on the red sequence. From our photometry, we have confirmed that our sample of passive spiral galaxy candidates do possess the multi-wavelength colours of passive galaxies. 

  \begin{table*}
  \caption{\textsc{galex}, SDSS and \textsc{\textit{WISE}} photometry of the passive spiral galaxy candidates in our sample (full table available online). We do not include $W4$ photometry here, as a large number of our galaxies do not have significant detections in this band.}
  \label{phot_table}
  \resizebox{\textwidth}{!}{
  \begin{tabular}{l c c c c c c c c c c c c c }
  \hline
  \textbf{Galaxy}   &\textbf{RA (J2000)} &\textbf{Dec (J2000)} &\textbf{T-type} & \textbf{$FUV$} & \textbf{$NUV$} & \textbf{$u$}   & \textbf{$g$}     & \textbf{$r$} & \textbf{$i$} & \textbf{$z$}    & \textbf{$W1$} & \textbf{$W2$} & \textbf{$W3$}\\ 
  \hline
NGC0015 & 2.26031 & 21.62449 &1& -- & 18.76 & 16.17 & 14.34 & 13.46 & 13.02 & 12.69 &12.99 &13.67 &14.61\\ 
 NGC0109 & 6.56101 &21.80736 &1& 19.17 & 18.66 & 15.74 & 14.16 & 13.35 & 12.91 & 12.61 &13.33 &13.97 &15.18\\ 
PGC001683 &6.83343 &-4.89138 & 5 & 20.89 & 19.63 & 16.22 & 14.44 & 13.64 & 13.21 & 12.94 &13.09 &13.77 &14.62\\ 
IC0022 &7.38824 &-9.08078 &3 & 21.22 & 19.34 & 16.24 & 14.44 & 13.64 & 13.24 & 12.96 &13.40 &14.06 & 15.53\\ 
  NGC0345 & 15.34209 &-6.88427 &1& 20.49 & 19.37 & 15.94 & 14.04 & 13.15 & 12.69 & 12.38 &12.75 &13.43 &14.97\\ 
  \hline
  \end{tabular}
  }
  \end{table*}

 \section{Spectroscopic Observations and Analysis}
 \label{observations1}
 In order to investigate whether emission lines are present despite a lack of photometric evidence of any star formation, we undertook a pilot program of IFS observations of six passive spiral galaxy candidates.
Due to observing scheduling constraints, we observed three galaxies with SDSS coverage and three southern-sky galaxies where DSS images were sufficient to visually confirm spiral morphology.

 Observations were taken on 2015 December 15th-18th  and 2016 March 11th-12th using the Wide Field Spectrograph \citep[WiFeS;][]{Dopita07, Dopita10} on the Australian National University's 2.3m telescope at Siding Spring Observatory. The WiFeS IFS has a $25\arcsec\times38\arcsec$ field of view, which corresponds to $\sim10 \times 16$ kpc at our mean redshift of $z\sim0.02$, and $1\arcsec\times1\arcsec$ spaxels (if $2\times1$ binning is employed in the y-direction) provide $\sim$0.4 kpc spatial resolution.
 The B3000 ($\sim$3500-5800~\AA) and R3000 ($\sim$5300-9000~\AA) gratings were used along with the RT560 dichroic, resulting in a spectral resolution of $\sigma\sim100~\textrm{km}~\textrm{s}^{-1}$ across the entire wavelength range. The instrument was used in nod-and-shuffle mode, with an average seeing of $1.8\arcsec$ throughout both runs. The data were reduced in the standard manner using the \textsc{Pywifes} reduction pipeline of \citet{Childress14}.
  \begin{table*}
  \caption{IFS observations of six passive spiral galaxy candidates taken 2015 December 15th-18th  and 2016 March 11th-12th. These six galaxies show absorption line spectra, with $\textrm{D}_{4000}$ values consistent with those of galaxies dominated by old stellar populations.}
  \label{observations}
  \begin{tabular}{l c c c c c c c}
  \hline
  \textbf{Galaxy}   & \textbf{RA}   &\textbf{Dec}     & \textbf{$z$}    &\textbf{Obs Date} & \textbf{Exp}        &\textbf{$\textrm{D}_{4000}^{1}$} & \textbf{Age}$^{2}$\\
                            & \textbf{(J2000)}       &   \textbf{(J2000)}       &                       & &\textbf{Time (s)}   &                                          &  \textbf{(Gyr)} \\
  \hline
  NGC 538   &          01:25:26.0  & -01:33:02 & 0.0183  &Dec 2015 & 3200   & 1.69 & 2.50\\
  IC 375       &           04:31:03.1  & -12:58:26 & 0.0351 &Dec 2015  & 4800  & 1.67 & 2.40\\
  MCG-02-12-043  & 04:35:11.1  & -13:14:40 & 0.0353 &Dec 2015  & 4500 & 1.54 & 1.43\\
  CGCG 394-006   & 04:46:25.6  & 00:21:59  & 0.0233 &Dec 2015  & 4200  & 1.80 & 3.25\\
  NGC 4305 &          12:22:03.6   & 12:44:27  & 0.0063  & Mar 2016 &   4500    &  1.50   & 1.02  \\
  NGC 4794 &          12:55:10.5 & -12:36:30  & 0.0132  & Mar 2016 &   2100  & 1.74  &    3.00 \\
  \hline
    \multicolumn{8}{l}{$^{1}$Using the narrow $\textrm{D}_{4000}$ definition of \citet{Balogh99}.}\\
  \multicolumn{8}{l}{$^{2}$Based on \citet{Bruzual03} models with Solar metallicity and a \citet{Chabrier03} IMF.}\\
  \end{tabular}
  \end{table*}

\begin{figure}
 \centering
  \includegraphics[width=3.5in]{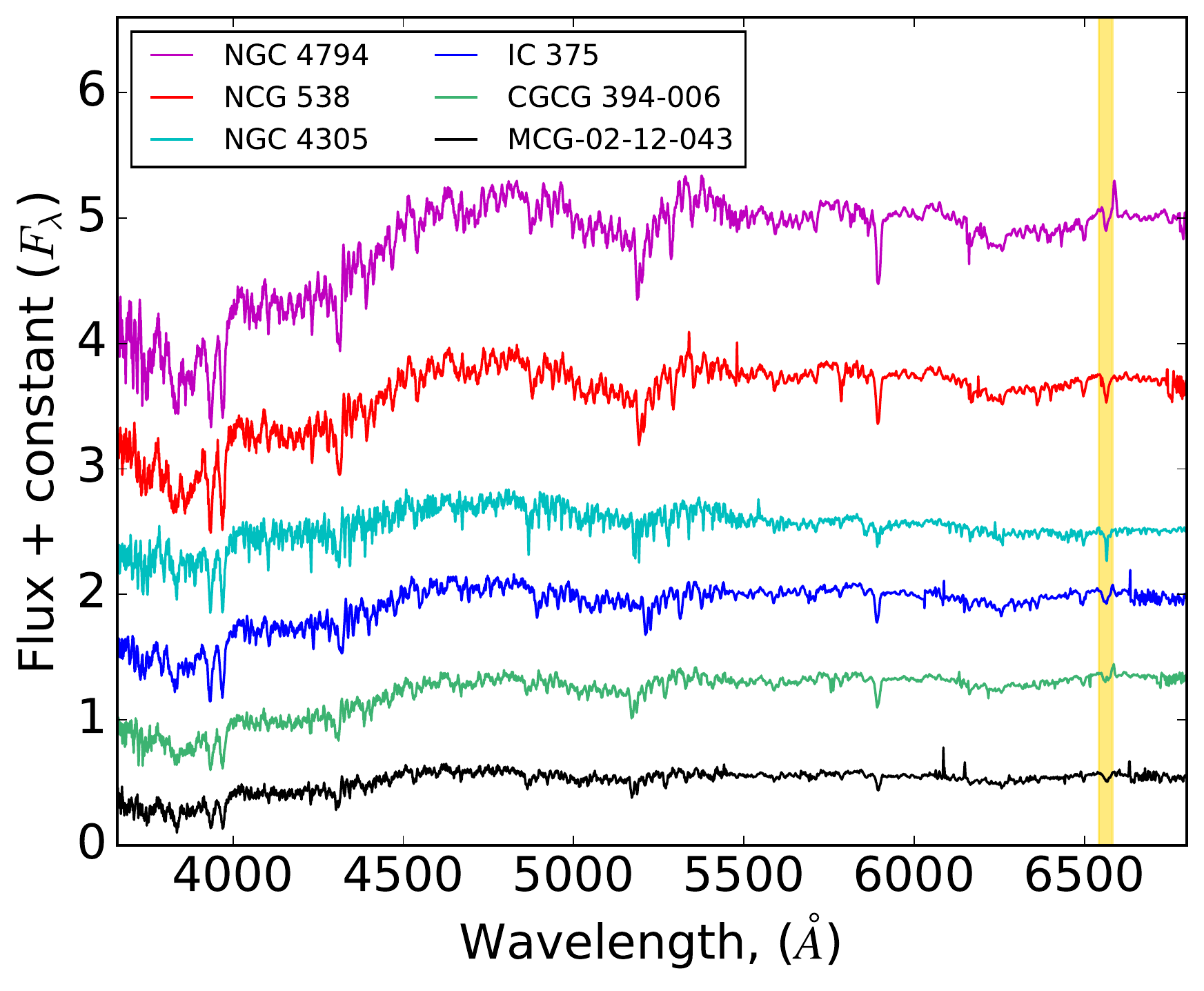}
  \caption{The integrated spectra of the six passive spiral galaxy candidates observed for this work, shifted to rest frame. H$\alpha$ regions are highlighted in yellow. Deep Balmer absorption lines are observed in all six galaxies, and moderate $D_{4000}$ breaks lead us to classify these spectra as being from ageing stellar populations with little or no star formation. }
  \label{Int_specs}
\end{figure}

  
  In Fig.~\ref{Int_specs} we plot the integrated spectra of the galaxies with the H$\alpha$ wavelength range highlighted in yellow. In each case, the data were Voronoi binned using the method of \citet{Cappellari03} to achieve adequate S:N per bin, and the stellar kinematics determined using the penalised pixel fitting method (\textsc{ppxf}) of \citet{Cappellari04}. To create the integrated spectrum, each Voronoi bin was velocity-shifted to that of the central spaxel, then stacked.\\
  All six galaxies show absorption line spectra, with deep Balmer lines. 
We fit and subtract a stellar continuum using \textsc{ppxf} for each integrated spectrum, and simultaneously measure any emission line flux by fitting single Gaussian profiles to emission line regions. To estimate the error on any H$\alpha$ detected, we perform 100 Monte Carlo realisations of the best fit to the integrated spectrum with random noise of the order of spectral variance added. An error measurement is calculated from the standard deviation of the subsequent H$\alpha$ measurements. For four of the six galaxies, the measured value of any H$\alpha$ is less than the estimated error in this measurement and we conclude their spectra are consistent with being passive. For the remaining two, IC 375 and CGCG 394-006, the H$\alpha$ measurement is small but significant, though even if we assume all H$\alpha$ is the result of star formation, in both cases the H$\alpha$-derived SFR is $<0.07~\textrm{M}\odot~\textrm{yr}^{-1}$ across the entire galaxy using the relation of \citet{Richards15}. 
From the nebular H$\alpha$ emission line measurements, we conclude there is little or no star formation within these galaxies. Further work to resolve spatial properties will show any radial variation across these galaxies.

  To confirm these galaxies are not post-starbursts, we measured H$\delta$ equivalent widths using the narrow definition of \citet{Worthey97}. Equivalent widths were $\textrm{H}\delta<1$\AA\ for all six galaxies, well below that required for a galaxy to be classified as post-starburst \citep[e.g.][]{Zabludoff96,Goto05}. This implies the bulk of star formation in these galaxies was not rapidly truncated in the last Gyr.
  
  We measured the 4000 \AA\ break ($D_{4000}$) of the integrated spectra following the procedure outlined by \citet{Balogh99}. The $D_{4000}$ values for all six galaxies, listed in Table~\ref{observations}, lie in the range of the \citet{Masters10} red spiral galaxies sample. 
We compared our $D_{4000}$ values to that of the single stellar population models of \citet{Bruzual03}. Using Padova 1994 evolutionary tracks with a \citet{Chabrier03} IMF and Solar metallicity we find the average luminosity-weighted age of our six red spirals is $\sim2.3$ Gyr, which is younger than that expected for a passive elliptical galaxy \citep[e.g.][]{Poggianti97}, though older than a star-forming spiral. The $D_{4000}$ break values (and hence the ages) are similar to the red spiral galaxies of \citet{Masters10}, so while our galaxies are passive now, they may have had some low level star formation recently.

   \section{Summary \& Conclusions}
   We have identified passive spiral galaxies in the nearby Universe and characterised their properties with multi-wavelength photometry and IFU spectroscopy. 
   Matched-aperture photometry in UV, optical and mid-IR wavebands of a sample of 51 galaxies is consistent a passive stellar population, and SDSS imaging confirming spiral morphologies. IFS of a pilot spectroscopic sample of six galaxies confirm an old stellar population dominates across both the disk and nuclear regions of the galaxies, with little or no H$\alpha$ emission. $D_{4000}$ values imply the average luminosity-weighted age for our spiral galaxies is $\sim2.3$ Gyr. Analysis of the kinematics, morphologies and environments of these galaxies will be presented in a forthcoming work.
  
Compared to previous samples of optically-red spiral galaxies, we have negligible contamination from star forming galaxies with H$\alpha$ nebular emission, mid-IR emission from warm dust nor UV emission from young stars.
Our sample has spiral morphologies, but multi-wavelength photometry and an absence of nebular emission are consistent with them being passive. We conclude a sample of passive spiral galaxies exist, and that quenching of star formation can occur without transformation into early-type morphologies.

\section*{Acknowledgements}
The authors would like to thank the referee whose comments have improved the quality of this manuscript.
AFM acknowledges the support of an Australian postgraduate award.
Funding for the SDSS III has been provided by
the Alfred P. Sloan Foundation, the U.S. Department of Energy Office of
Science, and the Participating Institutions.  This publication makes use of data products from the Wide-field Infrared Survey Explorer, which is a joint project of the University of California, Los Angeles, and the Jet Propulsion Laboratory/California Institute of Technology, funded by the National Aeronautics and Space Administration. Based on observations made with the NASA Galaxy Evolution Explorer. GALEX is operated for NASA by the California Institute of Technology under NASA contract NAS5-98034. The Digitized Sky Surveys were produced at the Space Telescope Science Institute under U.S. Government grant NAG W-2166, utilising photographic images obtained with the Oschin Schmidt Telescope on Palomar Mountain and the UK Schmidt Telescope.
  
    \bibliographystyle{mnras}
  \bibliography{RSbib}
    \end{document}